\documentclass{cupconf}
\usepackage{graphicx}

\begin{document}

\title[Chemical evolution]{Chemical Evolution models of Local Group galaxies}
\author[M. Tosi]{M\ls O\ls N\ls I\ls C\ls A T\ls O\ls S\ls I }
\affiliation{INAF - Osservatorio Astronomico di Bologna, Via Ranzani 1, I-40127
Bologna, Italy}

\maketitle

\begin{abstract}
 
{\it Status quo} and perspectives of standard chemical evolution models of Local
Group galaxies are summarized, discussing  what we have learnt from them, 
what we know we have not learnt yet, and  what I think we will learn in the 
near future. It is described how Galactic chemical evolution models have
helped showing that: 
i) stringent constraints on primordial nucleosynthesis 
can be derived from the observed Galactic abundances of the light elements, 
ii) the Milky Way has been accreting external gas from early epochs to the 
present time, 
iii) the vast majority of Galactic halo stars have formed quite rapidly at 
early epochs. 
Chemical evolution models for the closest dwarf galaxies, although still 
uncertain so far, are expected to become extremely reliable in the nearest 
future, thanks to the quality of new generation photometric and
spectroscopic data which are currently being acquired.

\end{abstract}

\firstsection
\section{Introduction}

The proximity of Local Group galaxies makes them the ideal benchmarks to study 
galaxy formation and evolution, because they are the only systems where the 
accuracy and the
wealth of observational data allows us to understand them in a sufficiently 
reliable way. In fact 
to understand the evolution of galaxies, astronomers must follow two distinct 
and complementary approaches: on the one hand they must develop theoretical 
models of galaxy formation, of chemical and  dynamical evolution, and, on 
the other hand, they must collect accurate 
observational data to constrain the models. Of particular 
importance is  to acquire reliable data on the chemical abundances,  masses 
and kinematics  of the galactic components (gas, stars, dark matter), on the 
star formation (SF) regimes, and on the stellar initial mass function (IMF),
quantities that are much better derivable in nearby systems.

Following Socrates' indication,
$\gamma\nu\hat{\omega}\theta\iota$  $\sigma\alpha\upsilon\tau${\it \'o}$\nu$ 
({\it know thyself}), that the knowledge of truth must be derived not from 
metaphysics but from critical analysis of the reality, here I try to critically
 review {\it status quo} and perspectives of
standard chemical evolution models, warning that this kind of models, although
quite successful, refer only to large-scale, long-term phenomena, and cannot
account for the small-scale, short-term variations often observed in the
chemical and dynamical properties of galaxies.

\section{Parameters}

The major parameters involved in standard chemical evolution models are:

\noindent$\bullet$
SF law and rate (often simplisticly approximated either as
exponentially decreasing functions of time, SFR $\propto e^{-t/\tau}$, or as
power laws of the gas density, e.g. SFR $\propto \Sigma_{gas}^n$);

\noindent$\bullet$ Gas flows in and out of the considered region (the infalling gas rate
being usually approximated with an exponentially decreasing function of time,
f$_i \propto e^{-t/\theta}$, and the galactic outflows, or winds, assumed to be
proportional to the energy released by Supernova explosions, f$_w \propto
E_{SN}$);

\noindent$\bullet$ IMF (usually represented as a power law, $\phi
\propto m^{-\alpha}$ with one or more exponents $\alpha$ for different mass
ranges, see Gallagher \& Grebel, this volume);

\noindent$\bullet$ Stellar lifetimes and nucleosynthesis yields.

\noindent Some of them, however, are implicitely linked to other
parameters, such as, for instance, the amount of mixing occurring in 
stellar interiors or the stellar mass loss rates. 

Since the parameters are many, the crucial prescription to avoid 
misleading
results from chemical evolution modeling is to {\bf always compare the model
predictions with all the available constraints}, not only with those relative to
the examined quantities. This prescription implies that, until now, Galactic
models are much better constrained than those for external, less studied
systems. In the following they are then discussed separately.

\section{The Galaxy}

In the case of the Galaxy, the observational constraints formally outnumber the
model parameters. Indeed, in the last two decades, an increasing number of
accurate and reliable data have been accumulated that allow us to put stringent
limits on the evolution of the Milky Way. The {\it minimal} list of data
that should always be  compared with the model predictions (see also
Boissier \& Prantzos 1999) includes:

\noindent$\bullet$
 current distribution with Galactocentric distance of the SFR (e.g. as
 compiled by Lacey \& Fall, 1985);
 
\noindent$\bullet$ current distribution with Galactocentric distance of the gas and star     
densities  (see e.g. Tosi, 1996, Boissier \& Prantzos 1999 and references therein);

\noindent$\bullet$ current distribution with Galactocentric distance of element abundances
 as derived from HII regions and from B-stars (e.g. Shaver et al. 1983,
 Smartt \& Rollerston 1997);
 
\noindent$\bullet$ distribution with Galactocentric distance of element abundances at
 slightly older epochs, as derived from PNe II (e.g. Pasquali \& Perinotto
 1993, Maciel \& Chiappini 1994, Maciel \& K\"oppen 1994, Maciel et al. 2003);
 
 \noindent$\bullet$ age-metallicity relation (AMR) not only in the solar neighbourhood but also
 at other distances from the center (e.g. Edvardsson et al. 1993);

 \noindent$\bullet$ metallicity distribution of G-dwarfs in the solar neighbourhood (e.g.
 Rocha-Pinto \& Maciel 1996);

\noindent$\bullet$  local Present-Day-Mass-Function (PDMF, e.g. Scalo 1986, Kroupa et al.
 1993);

\noindent$\bullet$  relative abundance ratios (e.g. [O/Fe] vs [Fe/H]) in disk and halo
 stars (e.g. Barbuy 1988, Edvardsson et al. 1993).

There are now several models able to reproduce all these observed properties.
Even if this circumstance does not provide yet a unique detailed scenario for 
the formation and evolution of the Milky Way, it allows us to make 
robust predictions on several important issues. 

\begin{figure}
\centerline{\includegraphics[width=12cm,height=7cm,clip]{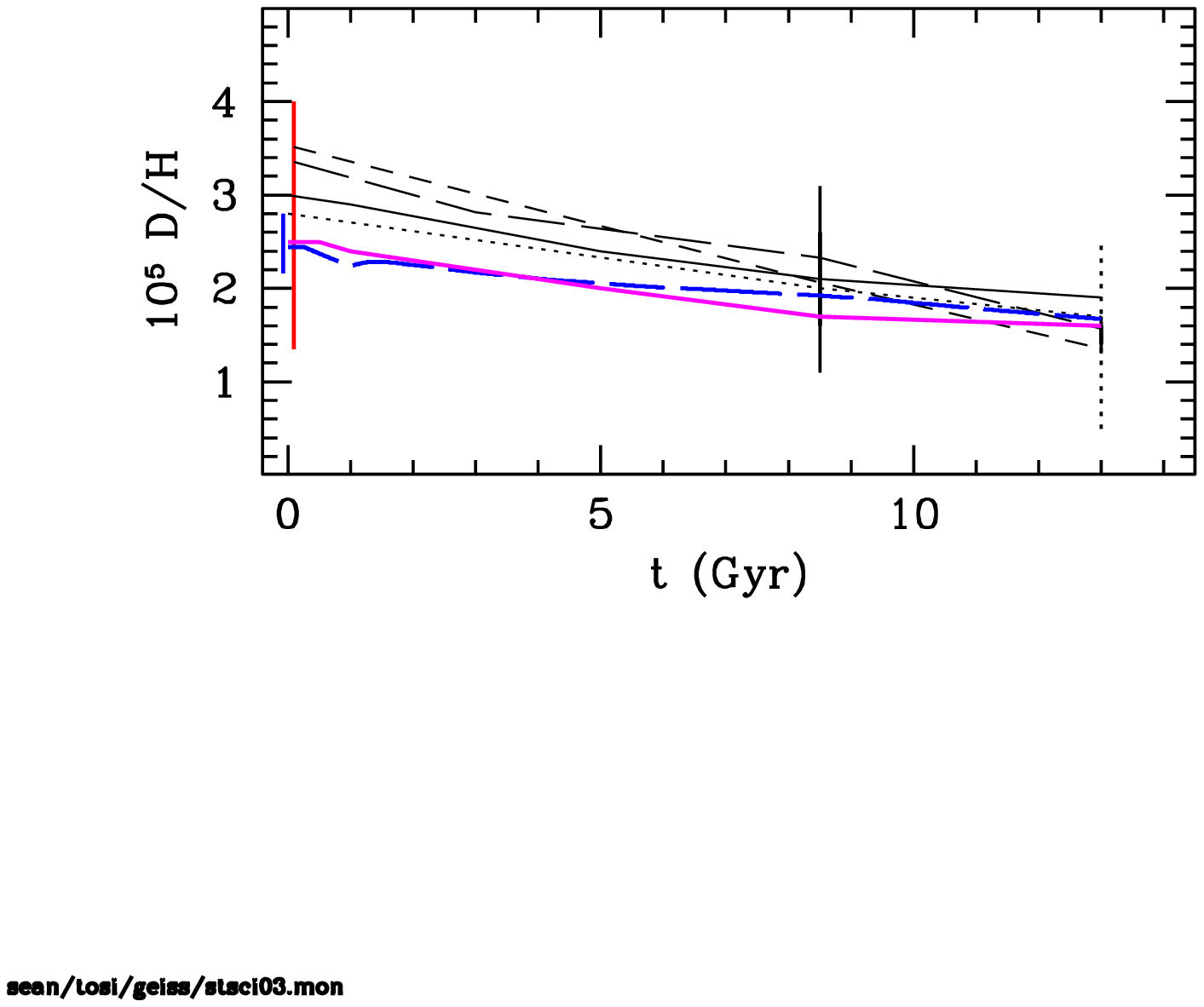}}
\caption{Deuterium evolution in the solar ring as predicted by various models:
  solid line, Steigman \& Tosi 1992; short-dashed line, Galli et al. (1995); 
  dotted line,  Prantzos (1996); long-dashed line, Chiappini et al. (1997). 
  The thick solid line and the thick long-dashed line are both from 
  Romano et al. 2003. 
  All the vertical lines correspond to data derived from observations; from left
  to right they are: the primordial D/H derived from WMAP, the range of
  abundances inferred from high-redshift QSO absorbers, the pre-solar value, and
  the local ISM value (see text for references). The dotted line at 13 Gyr shows
  the range of D/H ratios derived along different lines of sight (e.g.
  Vidal-Madjar et al. 1998, Sonneborn et al. 2000).
  }
\label{d}
\end{figure}

An example of robust prediction resulting from the requirement of
reproducing all the above list of data is the Galactic evolution of deuterium,
one of the elements produced during the Big Bang and one of the best
baryometers (see Steigman, this
volume).  When people began to study the 
evolution of D (e.g. Steigman \& Tosi 1992, Galli et al. 1995, Prantzos 1996), 
observational data on the D  abundance were 
available only for the local interstellar medium (ISM) and for the Protosolar 
Cloud (Linsky 1998, Geiss \& Gloeckler 1998 and references therein). 
All the models able to 
reproduce the above list of constraints (see Tosi 1996 and references therein) 
predicted only a 
moderate depletion of D from its primordial value to the present one: 
a factor of 3 at most (see Fig.\ref{d}). This implies a primordial number ratio 
to hydrogen 
(D/H)$_p \leq (4-5) \times 10^{-5}$ impossible to reconcile, within the framework of
standard Big Bang nucleosynthesis (SBBN), with the low primordial abundance by
mass of $^4$He, Y$_p \simeq 0.23$ inferred earlier on by several groups from low 
metallicity HII regions and globular clusters.
These Y$_p$ determinations have subsequently become the subjects of hot debates
(see Steigman, this volume, and references therein, for a critical discussion of
this result), but, at the time, this inconsistency led some
people think that the Galactic models were wrong and that we should find a way
to deplete much more D during the Galaxy evolution, to allow for a higher
primordial abundance. 
Then high-redshift, low-metallicity QSO absorbers started to be observed and 
provided D/H always lower than 4$ \times 10^{-5}$ (e.g. Burles \& Tytler 1998
and references therein), perfectly consistent with the
predictions of the Galactic models with low D depletion (see Fig.\ref{d}, Tosi
et al. 1998, Chiappini et al. 2003). However, concerns remained
that, despite their low-metallicity, high-redshift absorbers might have D 
contents lower than primordial, because some stellar activity could have 
already taken place there and burnt some of the original D. 
Eventually, a few months ago, the microwave satellite WMAP has provided a
direct estimate of the baryon-to-photon ratio, which corresponds, 
within the SBBN framework, to a primordial (D/H)$_p = (2.62 \pm 0.30) \times
10^{-5}$ (Spergel et al. 2003). This value is again in excellent agreement with
the predictions of Galactic chemical evolution models aimed at reproducing the
complete set of observational contraints (cfr. thick lines in Fig.\ref{d}, 
Romano et al. 2003) and
shows how robust model predictions could be when sufficiently constrained.
On the other hand, the significant variations of the D abundances measured in
the local ISM along different lines of sight (dotted vertical line in
Fig.\ref{d}) indicate that more sophisticated models would be needed to 
reproduce also local  fluctuations (see also Pilyugin \& Edmunds 1996).

 What have we learnt on the Galaxy formation and evolution from 
chemical evolution models ? One of the main results, in my opinion, is that
the Milky Way has not formed from a very rapid monolithic collapse of a single
proto-galaxy. We have recent observational evidences that the Galaxy is
accreting  the Sagittarius dwarf (e.g. Ibata et al. 1995), it is likely that it will accrete
the Magellanic Clouds in the future and someones (e.g.
Dinescu et al. 1999, Hilker \& Richtler 2000, Ferraro et al. 2002)
think that $\omega$Centauri is also
an accreted small dwarf, rather than a real globular cluster. In addition,
chemical evolution models already showed earlier on (e.g. Tinsley 1980, 
Tosi 1988 a, b, Matteucci \& Fran\c cois 1989, Chiappini et al. 1997, Boissier 
\& Prantzos 1999)
that the Milky Way must have kept accreting metal poor gas at a relatively 
steady rate.

\begin{figure}
\centerline{\includegraphics[width=12cm,height=9cm, clip]{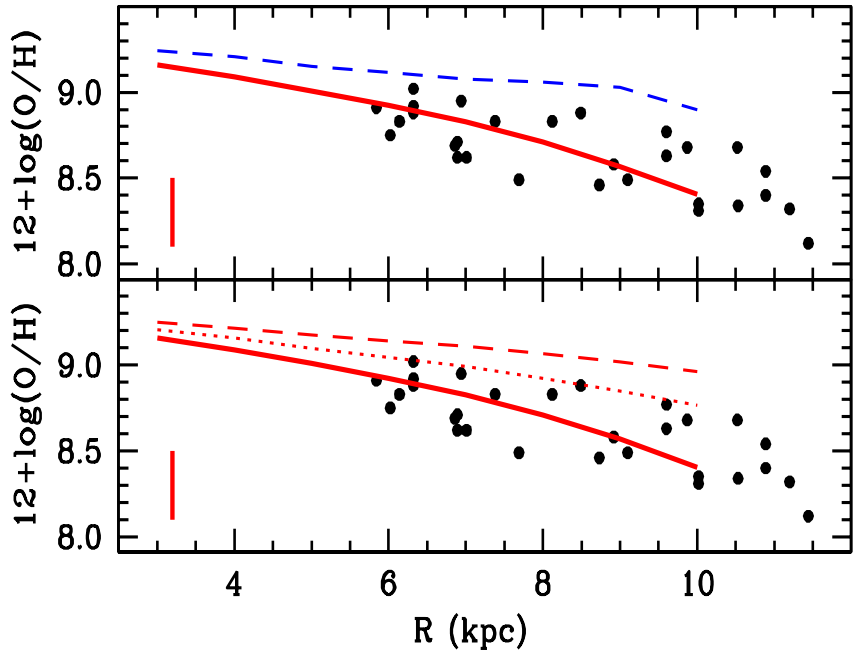}}
\caption{Radial distribution of the Galactic oxygen abundance at the present
 epoch, as derived from HII regions observations (dots) and as predicted by 
 chemical evolution models (Tosi 1988 a, b). 
 Top panel: models with primordial infall (solid
 line) and  without any infall after disk formation (dashed line). Bottom panel:
 models with infall of different metallicity (solid line for Z$_i$=0, dotted for 
 Z$_i$=1/2 Z$_{\odot}$, short-dashed for Z$_i$=Z$_{\odot}$). See text for 
 details. }
\label{grad}
\end{figure}

\begin{figure}
\centerline{\includegraphics[width=12cm, height=6cm,clip]{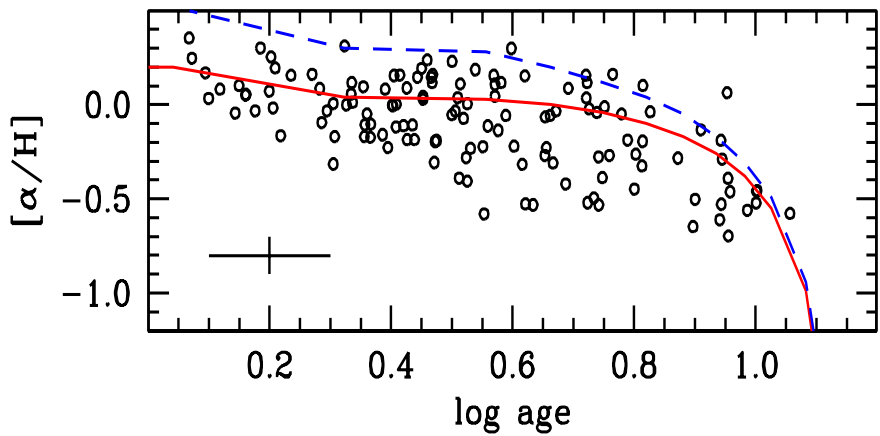}}
\caption{Age-metallicity relation in the solar neighbourhood as derived from
field star observations (dots) and as predicted by the models (lines) of the top 
panel of Fig.\ref{grad}. See text for details.}
\label{amr}
\end{figure}

Historically, one of the first reasons to invoke a continuous 
infall of metal poor gas on the Galactic disk
was the so-called G-dwarf problem (see Wyse, this volume), i.e.
the fact that, without infall, chemical evolution models overpredict the
number of low-metallicity long-lived stars. There are however other observed
properties that need infall to be reproduced. Figs \ref{grad} and \ref{amr} 
show two examples of this need. In Fig.\ref{grad} the radial distribution of the
current oxygen abundance in the disk is plotted as a function of Galactocentric
distance. In both panels the dots corresponds to HII regions values and the 
curves to the
predictions of Tosi (1988 a and b) models. When a constant equidense infall of
primordial gas is assumed (solid line in the top panel), the models reproduce
very well the observed distribution, whilst without any infall after the disk
formation (dashed line in the top panel) they 
predict an abundance gradient flatter than observed and 
overproduce oxygen at the present epoch at all galactocentric distances. 
The bottom panel of Fig.\ref{grad} illustrates why the accreted gas
should be metal poor: the solid line, as in the top panel, corresponds to a
metal free infall and is in excellent agreement with the data; the dashed line
corresponds to the same model, but assuming a solar infall metallicity, and
clearly overpredicts oxygen; the dotted line assumes a half-solar metallicity
and also overpredicts oxygen. 

Fig.\ref{amr} shows the age-metallicity relation in the solar neighbourhood, as
derived by Edvardsson et al. (1993) from observations of F and G dwarfs 
and as predicted
by the same models of the top panel of Fig.\ref{grad}. Again, the model 
assuming a constant infall of primordial gas (solid line) fits well the data, 
while the model with no infall after the disk formation (dashed line)
overproduces the metallicity since quite early epochs.
To reproduce both the observed radial gradients and abundances, 
nowadays all chemical
evolution models assume a fairly conspicuous amount of steady gas accretion
(but see also Lacey \& Fall 1985).
Most authors assume this gas to be primordial, but it was shown (Tosi 
1988b) that it could reach a metallicity
up to 0.2 Z$_{\odot}$ without loosing its diluting effects on the
predicted abundances. This metallicity happens to be that derived for the
few high-velocity clouds falling on the Galactic disk, 
where abundance measurements are possible (e.g. DeBoer \& Savage 1983).

\begin{figure}
\centerline{\includegraphics[width=12cm,height=9cm, clip]{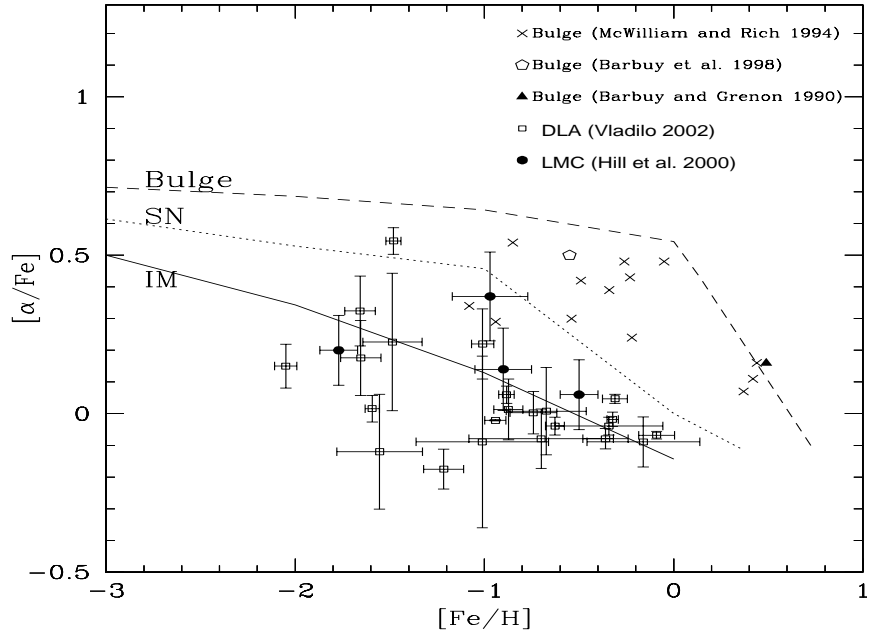}}
\caption{The lines show Matteucci's model predictions for 
the $\alpha$/Fe ratios for different SF regimes. See Matteucci (2003)
for references and details. }
\label{fran}
\end{figure}

\begin{figure}
\centerline{\includegraphics[width=12cm, height=9cm,clip]{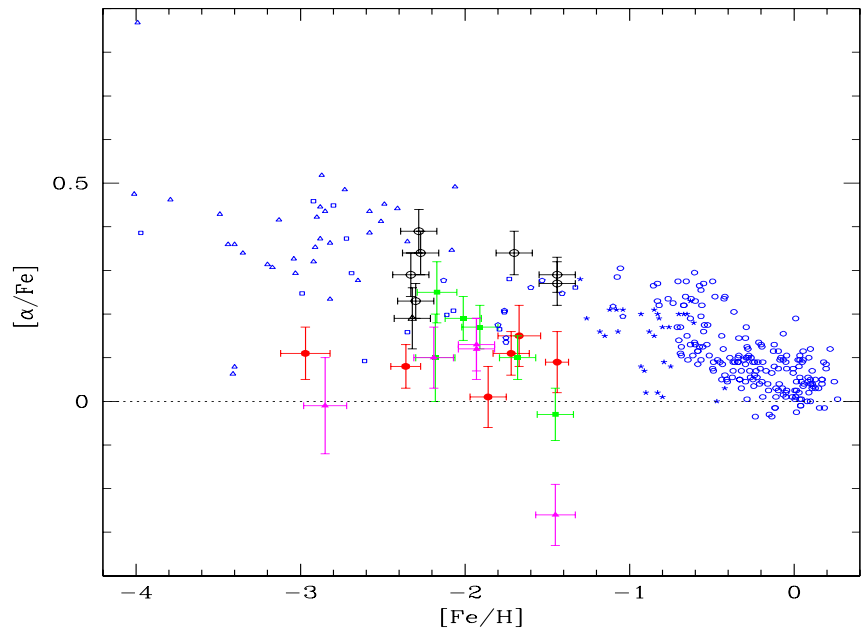}}
\caption{$\alpha$/Fe observed in different environments.
Open circles refer to local disk stars, open
triangles to local halo stars, open symbols with error bars to Galactic 
globular clusters and filled symbols with error bars to dSphs 
(Shetrone et al. 2001, and references therein).}
\label{shetr}
\end{figure}

Another important result is that 
both observations and chemical evolution models suggest that
the halo of the Galaxy must have created fairly rapidly most of its 
present stars. One of the main reasons to reach this conclusion comes from the
abundance ratios between alpha elements (those produced directly from $^4$He
burning, like oxygen and magnesium) and iron. 
Iron is mostly produced by Supernovae of type Ia (intermediate mass stars 
in binary systems) while alpha elements are synthesized
essentially in massive stars. The different lifetimes of their main producers 
imply that the enrichment of the alpha elements is very rapid (within a few
Myr) while that of iron starts to occur after $\sim$100 Myr and has its bulk
about 1 Gyr after the onset of the SF activity. As shown by
Matteucci (1992) this circumstance makes the stellar [$\alpha$/Fe]
measured in a region an excellent indicator of its SF regime. The curves in
Fig.\ref{fran} display  the different behaviours predicted by Matteucci's 
 models for different SF regimes. In regions,
 where the SF is supposed to be continuous over several
Gyrs, stars can form when iron has had plenty of time to enrich the medium
and contribute to their initial metallicity. Hence, the stellar 
[$\alpha$/Fe] is predicted to steadily decrease when iron increases (solid line
in Fig.\ref{fran}). Vice versa, in regions, like bulges, where the SF is 
very rapid and
intense, all the stars form prior to the release of iron by SNe Ia, and 
contain many $\alpha$ but little iron, thus showing a plateau (dashed
line) with high [$\alpha$/Fe]. In the solar neighbourhood, where some stars 
belong to the halo and others to the disk, Matteucci's models predict (dotted 
line) a fairly flat and high [$\alpha$/Fe] vs [Fe/H] for the halo stars with 
low [Fe/H] and a decreasing [$\alpha$/Fe] vs [Fe/H] for disk stars with higher
[Fe/H]. Since this is indeed the observed behaviour (see Fig.\ref{shetr}), 
this indicates that the halo must have had a strong and rapid initial SF
activity and the disk a continuous one.

Fig.\ref{shetr} shows [$\alpha$/Fe] vs [Fe/H] as derived from spectroscopy of
stars in various environments (Shetrone et al. 2001, and references therein). 
The small open circles correspond to solar
neighbourhood disk stars, the triangles to halo field stars, the larger open
circles with error bars to halo clusters, and they show an overall distribution
quite similar to that predicted by Matteucci's models. The filled symbols refer
instead to stars observed in nearby dwarf spheroidals (dSphs) and they all show 
[$\alpha$/Fe]  systematically lower than those measured at the same [Fe/H] in
Galactic stars. Further high-resolution spectroscopy of stars in other nearby
dSphs (Tolstoy et al 2003), including Sagittarius (Vladilo et al. this
conference and Bonifacio et al. 2003), have confirmed this systematic
difference. This evidence makes it extremely unlikely that our halo 
have formed mostly from the merging of dwarf galaxies like these, because there
is no conceivable mechanism able to make iron-poor, alpha-rich stars
assembling alpha-poor, iron-rich ones. 
From this and other arguments  (see Tosi 2003, and Wyse, this volume) it seems 
more
likely that our Galaxy has mainly formed from mostly gaseous building blocks and
within a relatively short timescale (a couple of Gyr, at most).

Despite the numeorus important achievements of chemical evolution modeling, 
there are important aspects of the Galaxy formation and evolution not well 
understood yet. Has the thin disk formed before or after the thick disk~? 
Where is the infalling gas coming from ? Is it actually 
as metal poor and as steady as required ?

A clear example of our lack of detailed 
knowledge of the processes leading to the
oberved Galactic properties is the evolution of the abundance gradients
in the disk. There are several chemical evolution models able to reproduce the
present metallicity gradients derived from HII regions and young stars
(e.g. Fig.\ref{grad}), along
with the whole set of constraints listed above. These models, however, differ
from each other in several assumptions and one of the major effects of such
differences is that they predict quite different evolutions of the gradient
(see e.g. Tosi 1996 and references therein): some predict the gradient to
steepen with time while others predict it to flatten. It is still difficult to
understand what the actual evolution is, because the available data refer
mostly to relatively young objects. Older single stars are in fact fainter and
hence more difficult to measure, specially at the large distances required to
derive the gradient with sufficient radial baseline. PNe in
principle are a good tool for this purpose, but the interpretation of their data
in terms of progenitor age is still rather uncertain and the progenitors
of the safest ones are stars a few Gyr old. The best targets to get a reliable
gradient back to the earliest epochs are open clusters, since they are much less
affected than individual objects by uncertainties on age, distance and
metallicity. Several people have derived the abundance gradient from open
cluster data (see e.g. Friel 1995 and references
therein), but what is needed for a robust result on the gradient 
evolution is a
cluster sample large enough to provide significant results in each age bin, and
with ages, metallicities and distances derived in a homogeneous way to avoid
spurious effects (e.g. Bragaglia et al. 2002). Building up such a sample 
clearly takes time, but we are confident that the results will be rewarding.

\section{Other Local Group galaxies}

The chemical evolution of M31 and M33 (as well as that of other spirals outside
the Group) has been modeled by a few groups with approaches similar to
those applied to the Milky Way (e.g. Diaz \& Tosi 1984, Moll\`a et al. 1996).
These models predict that the disks of these systems have a roughly continuous
SF, with shorter timescales for earlier type spirals and longer ones for later
type spirals. Infall of metal poor gas is required also for M31 (and other
massive spirals) whilst it is not necessary in low mass spirals like M33.
These models, however, are not as well constrained as those
for the Galaxy, since, at least until recently, the only data useful for
chemical evolution modeling were the HII region abundances and the gas and star
density distributions in the disks.

The data available for dwarf galaxies were equally scarse, but this kind of 
systems have intrigued many more scientists. In the last 25 years there has been 
a wealth of papers dealing with the chemical evolution of dwarfs (starting
e.g. with  Lequeux et al. 1979 and Matteucci \& Chiosi 1983): a rather
frustrating challenge, if one considers how inconsistent with each other the 
results of these papers have been. While most authors agreed that the IMF
in these galaxies is fairly similar to Salpeter's, both on the SF regimes and on
the existence of galactic winds different groups have reached very different
conclusions. For instance, many groups suggested that the SF in all dwarfs is
episodic (e.g. Matteucci \& Chiosi 1983, Pilyugin 1993, Larsen et al. 2001), but
others argue that in late-type dwarfs the SF is continuous (e.g. Carigi et al.
1995,  Legrand 2001). For the gas flows, some authors (e.g. Gilmore \& Wyse
1991)  concluded that winds triggered by SN explosions 
are not needed to explain the observed chemical abundances. However, other authors
reached the opposite conclusion that the winds are the only viable mechanism to
predict abundances as low as observed; some (e.g. Matteucci \& Tosi 1985,
Pilyugin 1993, Recchi et al. 2002) arguing that the outflowing gas must be
enriched in the elements produced by SNe and others (e.g. Pagel \&
Tautvaisiene 1998, Larsen et al. 2001) arguing, instead, that
the outflowing gas has the same composition as the galaxy medium.
In other words, all kinds of possible scenarios have been attributed to the
evolution of dwarf galaxies.

\begin{figure}
\centerline{\includegraphics[width=12cm, height=7cm,clip]{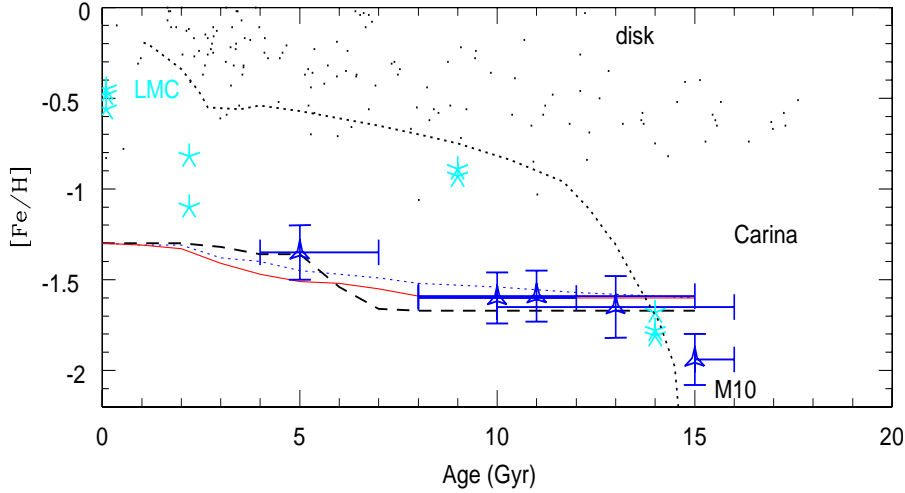}}
\caption{AMR derived from VLT/UVES spectroscopy of the dSph Carina
(triangles with error bars) by Tolstoy et al. (2003). Also shown for comparison
are the data relative to LMC clusters (asterisks), Pagel \& Tautvaisiene AMR
from LMC clusters (dotted line) and the data by Edvardsson et al. (1993) for
solar neighbourhood stars (dots).}
\label{tolstoy}
\end{figure}

\begin{figure}
\centerline{\includegraphics[width=13.5cm, height=8cm, clip]{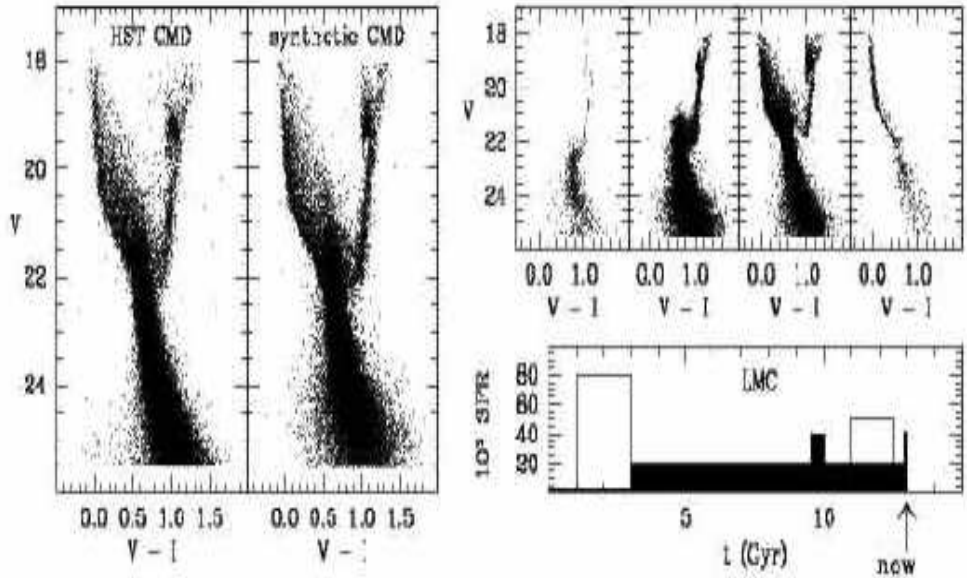}}
\caption{From the Coimbra experiment (Skillman \& Gallart 2002). 
  Left-hand panel: CMD of a
  field of the LMC bar derived by Smecker-Hane 
  et al. (2002) from HST/WFPC2 photometry; second panel from left: the
  corresponding best synthetic CMD by Tosi et al (2002). Top right panels:
  the synthetic CMD of the previous panel splitted in its four
  episode components, from the oldest to the youngest one from left to
  right. Bottom-right panel: the resulting SF history, i.e. SF rate per unit 
  area (in units of   10$^3$ M$_{\odot}$yr$^{-1}$kpc$^{-2}$) vs time is shown as filled
  histogram. The empty histogram refers to the SF history derived from LMC star
  clusters.
  }
\label{lmc}
\end{figure}

These inconsistencies are due to the lack of adequate observational data. 
Many groups, for instance, to model dwarf galaxy evolution have adopted 
the age-metallicity relation presented by Pagel \& Tautvaisiene 
(1998) for the LMC (dotted line in Fig.\ref{tolstoy}). 
However, the LMC is not a prototype for any kind of 
dwarfs, and the AMR was derived from clusters data, which do
not necessarily have the same evolution of field stars (see, in fact,
Fig.\ref{lmc}). I believe, however,
that the quantity and quality of the observational data on nearby dwarfs is  
so dramatically improving today that our understanding of dwarfs evolution
will make an impressive step forward in the near future. Indeed: 

1) High resolution spectrographs at 10 m class telescopes are providing 
in these months a wealth of new accurate abundances for field stars in nearby
dwarfs. The ages of these stars
can be derived from the colour-magnitude diagrams (CMDs) resulting from deep
high resolution photometry, both from ground and space.  
Fig.\ref{tolstoy} draws the first results by Tolstoy et al. 2003 from 
VLT/UVES observations and shows how different is the AMR of LMC clusters
and of Milky Way field stars from the AMR of the stars
in the Carina dwarf galaxy (as well as in the other
dwarfs of their program). 

2) Deep and tight CMDs provide reliable information on the IMF 
of the observed regions (see e.g. Gallagher \& Grebel, this volume) and on the
SF history of the observed regions.

The SF history is quite reliably derivable by
interpreting the observational CMDs with the synthetic CMD method (e.g. Tosi et
al. 1991), which has proven a powerful and robust tool, recently tested 
on an LMC field comparing the scenarios obtained by different groups 
(the so-called {\it
Coimbra experiment}, see Skillman \& Gallart 2002 and references
therein). Fig.\ref{lmc} shows in the bottom-right panel the SF history of the
field on the bar of the LMC as derived by our group (Tosi et al. 2002) 
applying the 
synthetic CMD method to the CMD obtained by Smecker-Hane et al. (2002) from
HST/WFPC2 photometry, and kindly provided for the Coimbra experiment. 
It is apparent that such beautiful data let them 
measure with sufficient accuracy even the oldest/faintest stars, thus allowing
for the derivation of the SF history back to the earliest epochs. The SF history
in this field of the LMC (filled histogram in Fig.\ref{lmc}) is fairly 
continuous, although with significant variations in the rate, and quite 
different from that inferred by Pagel \& Tautvaisiene (1998) from cluster
data (empty histogram). 

The combination of these high quality photometric and spectroscopic data, with
appropriate interpretation tools, will soon allow us to know the AMR, the SF
history and the IMF of nearby galaxies. Being the closest galaxies, the 
Magellanic Clouds are in the best position to allow for accurate and
extensive data sets on these quantities. Hence, 
HST photometry and 10 m class telescope spectroscopy can put us in a few years 
in the conditions of modeling their chemical evolution even more safely than 
that of the solar neighourhood. This opens new promising horizons to chemical 
evolution studies of dwarf galaxies in general and of late-type dwarfs in 
particular.  Taking into
account that the SMC can be considered a prototype for this kind of galaxies,
because it has their typical mass, gas fraction and metallicity, this would
imply an unprecedented step forward to understand the evolution of
galaxies that not only are the most numerous ones, but are also those that
can provide better clues to galaxy formation processes. This will confirm, once again,
that the Local Group is the best astrophysical laboratory to understand galaxy
evolution.       

\begin{acknowledgements}
I'm grateful to Johannes Geiss and 
all the colleagues of the ISSI LOLA--GE team for the enlightening discussions.
Mat Shetrone kindly provided his figure. 
This work has been partially supported by the Italian ASI and MIUR through    
grants  IR11301ZAM and Cofin-2002028935.
\end{acknowledgements}

\end{document}